\documentclass[12pt]{article}%
\usepackage{amsmath}
\usepackage{fancyhdr}
\usepackage{amsfonts}
\usepackage{amssymb}
\usepackage{graphicx}%
\begin{document}

\title{A different kind of quantum search}
\author{Lov K. Grover \thanks{Research was partly supported by NSA\ \&\ ARO under
contract DAAG55-98-C-0040.}
\and \textit{lkgrover@bell-labs.com}\\Bell Laboratories, Lucent Technologies,\\600-700 Mountain Avenue, Murray Hill, NJ 07974}
\date{}
\maketitle

\begin{abstract}
The quantum search algorithm consists of an alternating sequence of selective
inversions and diffusion type operations, as a result of which it can find a
target state in an unsorted database of size $N$ in only $\sqrt{N} $\ queries.
This paper shows that by replacing the selective inversions by selective phase
shifts of $\frac{\pi}{3},$ the algorithm gets transformed into something
similar to a classical search algorithm. Just like classical search algorithms
the algorithm has a fixed point in state-space toward which it preferentially
converges. In contrast, the quantum search algorithm moves\ uniformly in a
two-dimensional state space. This feature \ leads to robust search algorithms
and also to conceptually new schemes for error correction.

\end{abstract}

\section{Introduction}

\begin{quote}
The quantum search algorithm is like baking a souffle . \ . \ . . you have to
stop at just the right time or else it gets burnt \cite{brass_science}
\end{quote}

Search algorithms can be described as a rotation of the state vector in
2-dimensional Hilbert space defined by the initial and the target vectors. As
we describe later, any iterative quantum procedure \textit{has} to be a
continuos rotation in state space. In the original quantum search algorithm,
the state vector uniformly goes from the initial to the target and unless we
stop when it is right at the target, it will drift away. For many
applications, including unsorted database search, this leads to a square-root
speedup over the corresponding classical algorithm. One limitation of these
algorithms is that, to perform optimally, they need precise knowledge of
certain problem parameters, e.g. the number of target states.

In this paper we show that by replacing the selective phase inversions in
quantum search by suitable phase shifts we can get an algorithm that always
gives an improvement. As shown in figure 1, when a single iteration derived
from any unitary operator $U$ is applied, the state vector \textsl{always}
moves closer to the target state (Section 3). By recurring this basic
iteration, we develop an algorithm with multiple applications of $\mathit{U}$
that converge monotonically to the target (Section 4). This leads to variants
of quantum searching that are robust to changes in the parameters (Section 5).
Also, this immediately leads to schemes for reducing certain kinds of errors
in quantum computing (Section 6).%
\begin{figure}
[ptb]
\begin{center}
\includegraphics[
trim=0.000000in 0.887953in 0.000000in 2.868597in,
height=1.337in,
width=3.5276in
]%
{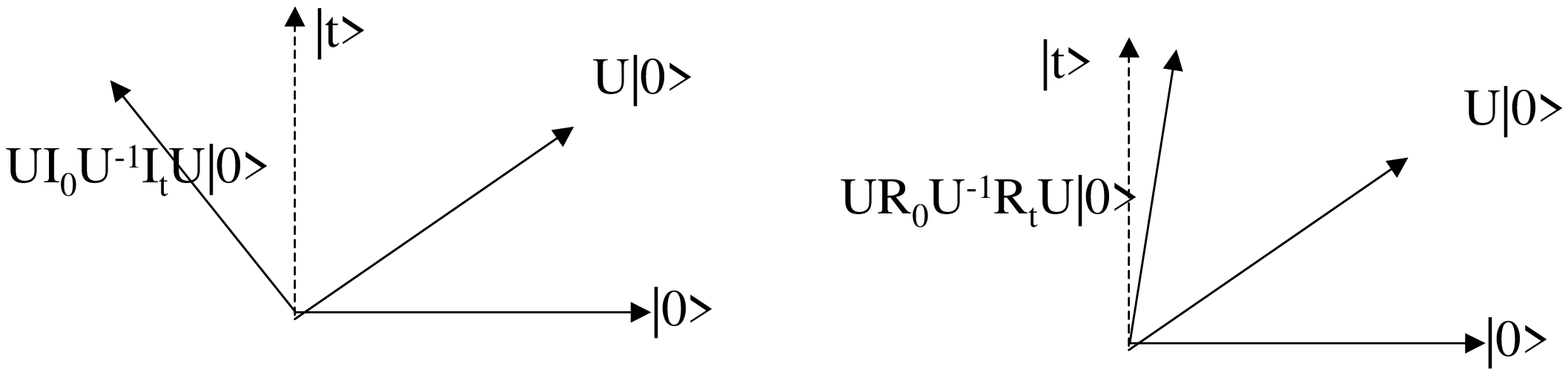}%
\caption{In the quantum search algorithm (left), the state vector overshoots
the target state; in the algorithm of this paper (right), the state vector
always moves towards the target.}%
\end{center}
\end{figure}

\section{A different kind of quantum search}

Consider the following transformation%

\begin{equation}
UR_{s}U^{\dagger}R_{t}U\left\vert s\right\rangle \label{transf}%
\end{equation}
$R_{t}$ \& $R_{s}$ denote selective phase shifts of the respective state(s) by
$\frac{\pi}{3}.$ Note that if we were to change these phase shifts from
$\frac{\pi}{3}$ to $\pi,$\ we would get one iteration of the amplitude
amplification algorithm \cite{bht}, \cite{ampt. amp.}.

The next section shows that if the $U$ operation drives the state vector from
a source $\left(  s\right)  $ to a target $\left(  t\right)  $ state with a
probability of $\left(  1-\epsilon\right)  $, i.e. $\left\Vert U_{ts}%
\right\Vert ^{2}=\left(  1-\epsilon\right)  $, then the transformation
(\ref{transf}) drives the state vector from the source to the same target
state with a probability of $\left(  1-\epsilon^{3}\right)  .$ The deviation
from the $t$ state has hence fallen from $\epsilon$\ to $\epsilon^{3}$.

The striking aspect of this result is that it holds for \textit{any} kind of
deviation from the $t$\ state. Unlike the standard amplitude amplification
algorithm which would overshoot the target state when $\epsilon$ is small
(Figure 1); the new algorithm will always move towards the target. As shown in
Section 5, this can be used to develop algorithms that are more robust to
variations in the problem parameters.

Connections to error correction might already be evident in the previous
paragraph. Let us say that we are trying to drive a system from an $s$\ to a
$t$\ state/subspace. The transformation that we have available for this is $U
$ which drives it from $s$\ to $t$\ with a probability $\left\Vert
U_{ts}\right\Vert ^{2}$ of $\left(  1-\epsilon\right)  ,$ i.e. the probability
of error in this transformation is $\epsilon$. Then the composite
transformation $UR_{s}U^{\dagger}R_{t}U$\ will reduce the error to
$\epsilon^{3}.$

This technique is applicable whenever the transformations $U,\ U^{\dagger
},R_{s}\ \&\ R_{t}$ can be implemented. This will be the case when errors are
systematic errors or slowly varying errors, e.g. due to environmental
degradation of some component. This would not apply to errors that come about
as a result of sudden disturbances from the environment. It is further assumed
that the transformation $U$ can be inverted with exactly the same error
(illustrated in Section 6). Traditionally quantum error correction is carried
out at the single qubit level where individual errors are corrected, each
error being corrected in a separate way. With the machinery of this paper,
errors can be corrected without ever needing to identify the error syndrome.

\section{Analysis}

We analyze the effect of the transformation $UR_{s}U^{\dagger}R_{t}U$ when it
is applied to the $\left\vert s\right\rangle $ state. As mentioned in the
previous section, $R_{t}$ \& $R_{s}$ denote selective phase shifts of the
respective state(s) by $\frac{\pi}{3}$ ($t$ for target, $s$ for source). We
show that if $\left\Vert U_{ts}\right\Vert ^{2}=\left(  1-\epsilon\right)  ,$
then
\[
\left\Vert \left\langle t\right\vert UR_{s}U^{\dagger}R_{t}U\left\vert
s\right\rangle \right\Vert ^{2}=\left(  1-\epsilon^{3}\right)  .
\]

In the rest of this section, the greek alphabet $\theta$ will be used to
denote $\frac{\pi}{3}.$ Start with $\left\vert s\right\rangle $ and apply the
operations $U,R_{s},U^{\dagger},R_{t}$ \&$\ U.$ If we analyze the effect of
the operations, one by one, just as in the original quantum search algorithm
\cite{grover96}, we find that it leads to the following superposition:%
\[
U\left\vert s\right\rangle \left(  e^{i\theta}+\left\Vert U_{ts}\right\Vert
^{2}\left(  e^{i\theta}-1\right)  ^{2}\right)  +\left\vert t\right\rangle
U_{ts}\left(  e^{i\theta}-1\right)  .
\]

To estimate the deviation of this superposition from $\left\vert
t\right\rangle $, consider the amplitude of the above superposition in
non-target states. The probability is given by the absolute square of the
corresponding amplitude:%
\[
\left(  1-\left\Vert U_{ts}\right\Vert ^{2}\right)  \left\Vert \left(
e^{i\theta}+\left\Vert U_{ts}\right\Vert ^{2}\left(  e^{i\theta}-1\right)
^{2}\right)  \right\Vert ^{2}.
\]

Substituting $\left\Vert U_{ts}\right\Vert ^{2}=\left(  1-\epsilon\right)  ,$
the above quantity becomes$:$%
\begin{align*}
&  \epsilon\left\Vert \left(  e^{i\theta}+\left(  1-\epsilon\right)  \left(
e^{i\theta}-1\right)  ^{2}\right)  \right\Vert ^{2}\\
&  =\epsilon\left\Vert \left(  -e^{i\theta}+e^{2i\theta}+1\right)
-\epsilon\left(  e^{i\theta}-1\right)  ^{2}\right\Vert ^{2}\\
&  =\epsilon^{3}.
\end{align*}
The following sections give two simple applications of the above analysis -
the first to searching in the presence of uncertainty and the second to error correction.

\section{Recursion}

A few years after the invention of the quantum search algorithm
\cite{grover96}, \cite{sch} it was generalized to a much larger class of
applications known as the amplitude amplification algorithms \cite{bht},
\cite{ampt. amp.}. In these algorithms, the amplitude produced in a particular
state $t$ by starting from a state $s$\ and applying a unitary operation $U$,
can be \textit{amplified} by successively repeating the sequence of
operations:\ $Q=I_{s}U^{\dag}I_{t}U$. Here $I_{s}\ \&\ I_{t}$ denote selective
inversions of the $s\ \&\ t$ states respectively. For later reference, note
that the amplitude amplification transformation with four queries is:%
\begin{equation}
U\left(  I_{s}U^{\dagger}I_{t}U\right)  \left(  I_{s}U^{\dagger}I_{t}U\right)
\left(  I_{s}U^{\dagger}I_{t}U\right)  \left(  I_{s}U^{\dagger}I_{t}U\right)
\label{ampt-amp}%
\end{equation}
If we start from the $s$ state and repeat the operation sequence $I_{s}%
U^{\dag}I_{t}U,$ $\eta$ times, then the amplitude in the $U^{\dag}\left\vert
t\right\rangle $ state becomes approximately 2$\eta U_{ts}$ provided $\eta
U_{ts}\ll1$. The quantum search algorithm is a particular case of amplitude
amplification with $U$ being the Walsh-Hadamard Transformation ($W$) and $s$
being the $\overline{0}$ state (state of the system with all qubits in the $0$
state). The selective inversions enable the amplitudes produced in the various
iterations to add up in phase. The amount of amplification increases linearly
with the number of repetitions of $Q$ and hence the probability of detecting
$t$ goes up quadratically.

Just like the amplitude amplification transformation, it is possible to recurs
the transformation $UR_{s}U^{\dagger}R_{t}U$\ $\left\vert s\right\rangle $ to
obtain larger rotations of the state vector in a carefully-defined two
dimensional Hilbert space. \ This recursion will be described in detail in
\cite{tath}, the basic idea is to define transformations $U_{m}$ by the
recursion:
\begin{equation}
U_{m+1}=U_{m}R_{s}U_{m}^{\dagger}R_{t}U_{m},\ \ \ \ \ \ \ \ U_{0}=U.
\label{algorithm}%
\end{equation}
Unlike amplitude amplification, it is \textit{not} simple to write down the
precise operation sequence for $U_{m}$ with large $m$ without working out the
full recursion for all integers less than $m$. Recursion for each $m$ is
different and there is no simple structure. Let us illustrate this for $U_{2}$:%

\[
U_{0}=U
\]%
\[
U_{1}=U_{0}R_{s}U_{0}^{\dagger}R_{t}U_{0}=UR_{s}U^{\dagger}R_{t}U
\]%
\begin{align}
U_{2}  &  =U_{1}R_{s}U_{1}^{\dagger}R_{t}U_{1}=\left(  UR_{s}U^{\dagger}%
R_{t}U\right)  R_{s}\left(  UR_{s}U^{\dagger}R_{t}U\right)  ^{\dagger}%
R_{t}\left(  UR_{s}U^{\dagger}R_{t}U\right) \nonumber\\
&  =\left(  UR_{s}U^{\dagger}R_{t}U\right)  R_{s}\left(  U^{\dagger}%
R_{t}^{\dagger}UR_{s}^{\dagger}U^{\dagger}\right)  R_{t}\left(  UR_{s}%
U^{\dagger}R_{t}U\right) \nonumber\\
&  =U\left(  R_{s}U^{\dagger}R_{t}U\right)  \left(  R_{s}U^{\dagger}%
R_{t}^{\dagger}U\right)  \left(  R_{s}^{\dagger}U^{\dagger}R_{t}U\right)
\left(  R_{s}U^{\dagger}R_{t}U\right)  \label{recursion3}%
\end{align}
The corresponding transformation for amplitude amplification is
(\ \ref{ampt-amp}).

It is straightforward to show that if $\left\Vert U_{ts}\right\Vert
^{2}=1-\epsilon$, then $\left\Vert U_{m,ts}\right\Vert ^{2}=1-\epsilon^{3^{m}%
}$. Expressed as a function of the number of queries ($q_{m}$) $\left\Vert
U_{m,ts}\right\Vert ^{2}=1-\epsilon^{2q_{m}+1}.$The failure probability hence
falls as $\epsilon^{2q_{m}+1}$ after $q_{m}$ \ queries \cite{tath}; this is
similar to a classical algorithm where the probability of failure falls as
$\epsilon^{q+1}$ after $q$ queries (a classical algorithm is discussed in
Section 5).

\subsection{Fixed point of algorithm}

First, note that the standard amplitude amplification algorithm
(\ref{ampt-amp}) and the phase shift algorithm (\ref{recursion3}), both have
some selective operations performed on the t-state and so from an information
theoretic point of view there is no violation in having fixed points. However,
unitarity would be violated if there was any kind of accumulation at the
target state due to repetition of the same transformation. In amplitude
amplification (\ref{ampt-amp}), exactly the same transformation is repeated
and so unitarity does not permit any fixed point. In the phase shift algorithm
(\ref{recursion3}), which is very similar to amplitude amplification, the
transformation repeated in each step is slightly different due to the presence
of each of the four operations $R_{s},R_{t},R_{s}^{\dagger},R_{t}^{\dagger}$
and it hence gets around the unitarity condition that prevents amplitude
amplification from having a fixed point.

\section{Quantum searching amidst uncertainty}

The original quantum search algorithm is known to be the best possible
algorithm for exhaustive searching \cite{bbbv}, \cite{zalka} therefore no
algorithm will be able to improve its performance. However, for applications
other than exhaustive searching for a single item, this paper demonstrates
that suitably modified algorithms may indeed provide better performance.

Consider the situation where a large fraction of the states are marked, but
the precise fraction of marked states is not known. The goal is to find a
single marked state with as high a probability as possible in a single query.
For concreteness, say some unknown fraction, $f$ , of the states are marked,
with $f$ uniformly distributed \ between 75\% and 100\% with equal probability.

In the following we show that the probability of failure for the new scheme is
approximately one fourth that of the best (possible) classical scheme. Also,
it is approximately one fourth of that of the best (known) quantum scheme.

\paragraph{Classical}

The best classical algorithm is to select a random state and see if it is a
$t$\ state (one query). If yes, return this state; if not, pick another random
state and return that. The probability of failure is equal to that of not
getting a single $t$\ state in two random picks, i.e. $(1-f)^{2}$ which lies
in the range $\left(  0,0.06\right)  $. The overall failure probability is
approximately 3.12$\%$.
\begin{figure}
[ptb]
\begin{center}
\includegraphics[
trim=0.977980in 1.958896in 0.000000in 1.861401in,
height=1.9121in,
width=4.6484in
]%
{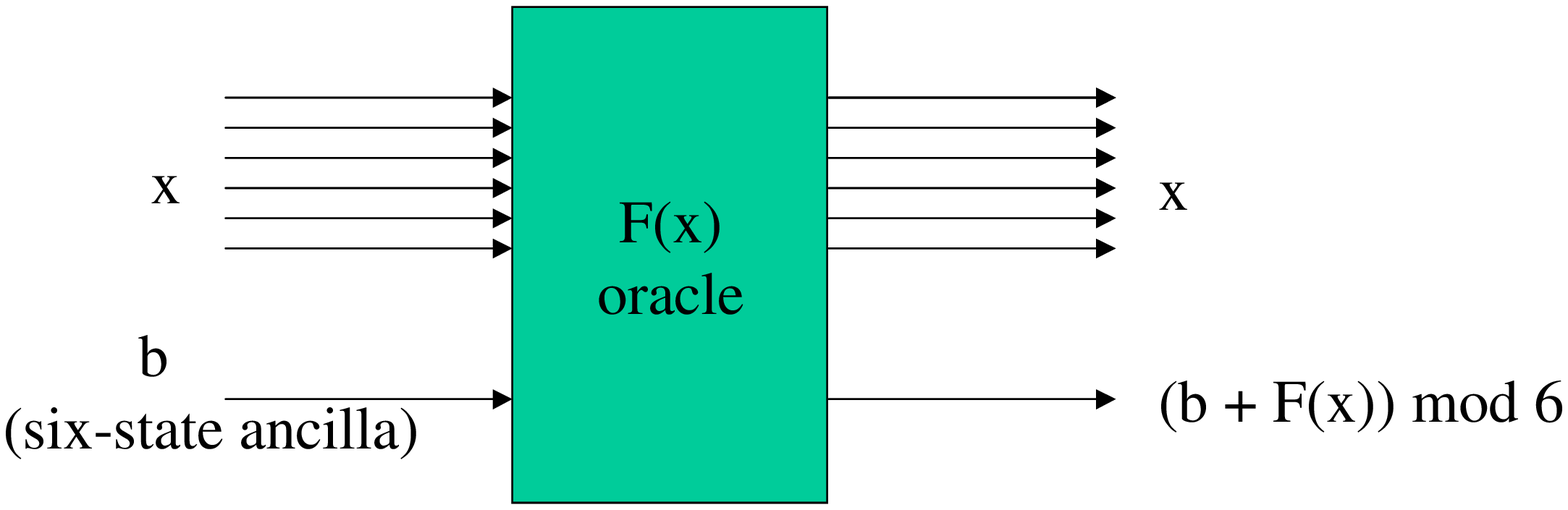}%
\caption{By setting the six-state ancilla, b, to the superposition $\frac
{1}{\sqrt{6}}\left(  \left\vert 0\right\rangle +\left\vert 1\right\rangle
\omega+\left\vert 2\right\rangle \omega^{2}+\left\vert 3\right\rangle
\omega^{3}+\left\vert 4\right\rangle \omega^{4}+\left\vert 5\right\rangle
\omega^{5}\right)  $ where $\omega=\exp\left(  -\frac{i\pi}{3}\right)  ,$ we
get a $\frac{\pi}{3}$ phase-shift of the states for which $F(x)=1$ relative to
those for which $F\left(  x\right)  =0.$ A simpler implementation using binary
qubits is presented in \cite{tath}.}%
\end{center}
\end{figure}

\paragraph{Quantum Searching}

The best quantum search based algorithm for this problem that I could find in
the literature was by Ahmed Younes et al \cite{younes}. This finds a solution
with a probability of $\left(  1-\cos\theta\right)  \left(  \frac{\sin
^{2}\left(  q+1\right)  \theta}{\sin^{2}\theta}+\frac{\sin^{2}q\theta}%
{\sin^{2}\theta}\right)  ,$ where $q$ = number of queries and $\theta
=\arccos(1-f)$ (Equation (59) from \cite{younes})$.$ When $q=1,$ the success
probability becomes: $f\left(  1+4(1-f)^{2}\right)  ,$ this lies in the range:
$\left(  0.94,1\right)  $. The overall failure probability is approximately
3.12$\%$.

\paragraph{New algorithm}

If we apply the phase shift transformation $UR_{s}U^{\dagger}R_{t}%
U$\ $\left\vert s\right\rangle $ (\ref{transf}) with $U$\ being the W-H
transform ($W$) and the state $s$\ being the $\overline{0}$ state (state with
all qubits in the 0 state), then $\left\Vert U_{ts}\right\Vert ^{2}=f,$ where
$f$ lies in the range $\left(  0.75,1.0\right)  .$ In the terminology of the
previous section, $\epsilon$ is defined by the equation, $\left\Vert
U_{ts}\right\Vert ^{2}=1-\epsilon.$ Therefore $\epsilon=1-f$ and after the
transformation $WR_{s}WR_{t}W\left\vert 0\right\rangle $, the probability of
being in a non-$t$\ state becomes $\epsilon^{3}$ which is equal to
\ $(1-f)^{3}$, i.e. the chance of a failure lies in the range $(0,0.0016)$.
The overall failure probability is approximately 0.8$\%.$%
\begin{figure}
[ptb]
\begin{center}
\includegraphics[
trim=0.000000in 0.758960in 0.000000in 0.688463in,
height=2.2684in,
width=3.7273in
]%
{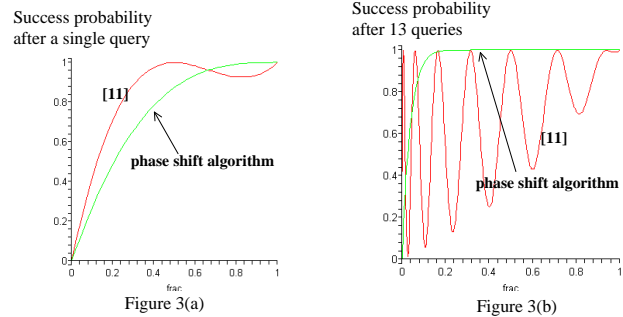}%
\caption{Comparison of the performance of the $\pi/3$ phase shift algorithm
with \cite{younes}, when the fraction of marked states ($f)$, varies between 0
\& 1.}%
\end{center}
\end{figure}

The performance of the algorithm is graphically illustrated in figure 3(a). In
the region of interest of this problem, the graph of the phase shift algorithm
lies entirely above that of the graph of \cite{younes} \ \ everywhere and so
the averaged success probability of the phase shift algorithm will clearly be
higher. The difference in the two becomes even more dramatic if we consider
multiple query algorithms (Figure 3(b)).

In \cite{pi_3_srch}, it will be shown that the phase-shift algorithm of this
paper, for the type of problems discussed in this section, is the best
possible quantum algorithm asymptotically.

\section{Quantum Error Correction}

\subsection{Background}

Von Neumann observed in 1944 that if a certain module had an error probability
of $\epsilon,$ then the error probability\ due to this module can be reduced
by doing the computation just three times and \ then deciding which state
occurs most often in the output \cite{Von56}. Assuming a perfect vote, the
error probability of the modified scheme is $O\left(  \epsilon^{2}\right)  .$
The approach of Von Neumann is still intact - add a small amount of redundancy
to the circuit by means of which one can infer whether or not the solution is
correct. If incorrect, redo the computation. However, in quantum circuits,
this approach does not work in the above form due to the different nature of
quantum information. It is not possible to observe quantum information without
affecting it.

Remarkably, in 1996 Peter Shor \&\ Andrew Steane \cite{ChNi} independently
discovered that it was possible to correct small errors in quantum information
even within the limitation of the above rules. They both did this by encoding
each qubit into multiple qubits in a way that as long as the error affected
any single qubit, it got projected into an orthogonal subspace where it could
be identified and corrected. So the principle was established - quantum error
correction was possible.

Unfortunately, the error correction provisions are very demanding and
considerably increase the complexity of the circuit. There have been several
schemes proposed for quantum error correction.\ Most schemes have the
limitation that they require the error per gate to be very small (of the order
of $10^{-4}$) and/or require a large number of gates. This paper presents a
new scheme based on the quantum search algorithm that can be used in
conjunction with other schemes to reduce systematic errors.

Let us say that we want to implement a certain transformation $U$ to drive the
system into a $t$\ state (or subspace) with certainty. However, when $U$ is
applied to the starting state s, the probability of reaching $t$\ is only
$\left(  1-\epsilon\right)  $, i.e. $U$ produces an error of $\epsilon$. Just
like Von Neumann had observed for classical circuits, we show that by doing
the transformation $U$ three times, we can considerably reduce the error.
However, the similarity ends there - the implementation and the error
correction technique is very different from classical.

The analysis of Section 3 shows that if we can apply the composite operation
$UR_{s}U^{\dagger}R_{t}U\left\vert s\right\rangle $, then, by a suitable
choice of $s~\&~\ U,$ we can reduce the error from $\epsilon$\ to
$\epsilon^{3}.$ This implementation thus depends on our ability to efficiently
apply the operations $U,\ R_{s},\ U^{\dagger}\ \&\ R_{t}.$

The operation $U$\ is the one being corrected and we assume that we can apply
it two times just as easily. Since quantum gates are reversible, we assume
that we can also apply $U^{\dagger}$ as easily (note that this must reuse the
same or very similar hardware as what $U$ did so as to keep the error exactly
the same). For systematic errors and slowly varying random errors, this can
probably be achieved since we may assume that the circuit parameters stay
fixed in time.

$R_{s}\ \&\ R_{t}$ require us to selectively shift the phases of certain
states. Shifting the phase of a state is as easy as identifying the presence
of the state (Figure 2). This leads to a number of different error-correction
schemes, depending on the type of error to be corrected.

To summarize, the error-correction technique requires the following conditions
to be satisfied:

\begin{enumerate}
\item In case we are correcting errors in a transformation, $U$, we should be
able to apply $U$\ twice and $U^{\dagger}$ once. These transformations must be
applied with exactly the same error as in the original $U$.

\item We should have a sub-module to distinguish the signal part of the output
wavefunction from the error. This is necessary to carry out $R_{t}$.

\item Finally, we assume the ability to perform noiseless $R_{t}$ \& $R_{s}$ operations.
\end{enumerate}

The forthcoming paper \cite{reich}, shows in detail how the methodology of
this paper can be used to design elementary (one \&\ two qubit gates) that
perform precisely even in the presence of small errors in $R_{s}\ \&\ R_{t}$.

\subsection{Example - Communicating Classical Bits}

We illustrate this error-correction procedure with a simple example. Consider
the problem of transmitting classical information over a quantum channel.
Although the channel is quantum, the information of interest is classical.
Therefore the only portion of the errors that are of concern are the amplitude
errors (i.e. bit-flip errors), we do not care about the phase. It is
well-known that by adding a single parity bit, we will be able to identify the
presence of single bit-flip errors. To correct these would normally require
additional bits. By making use of the error correction scheme of this paper,
we show how to \textit{correct} single bit-flip errors using just a single
parity qubit.
\begin{center}
\includegraphics[
trim=0.000000in 1.370927in 0.000000in 1.648412in,
height=2.6195in,
width=5.8176in
]%
{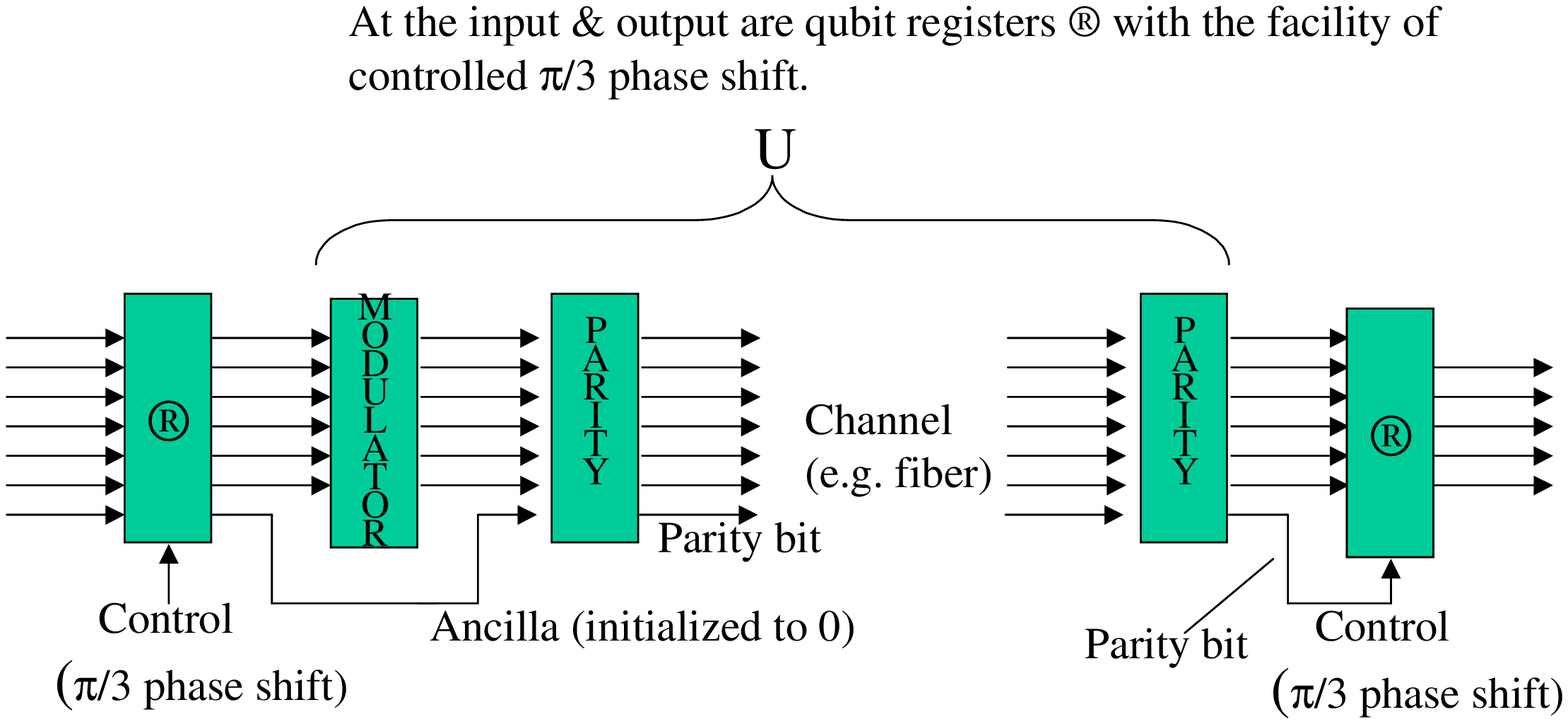}%
\\
Figure 4 - Redundancy, in the form of a parity bit, helps to detect, and
\textit{correct}, single bit-flip errors.
\end{center}

\subsubsection{Building blocks}

\begin{enumerate}
\item The input and output registers have a provision for conditionally
shifting the phase of the state of qubits by $\frac{\pi}{3}.$

\item The modulator flips the state of certain qubits depending on the message
to be transmitted.

\item There are two parity generators that take as input ($\eta$+1) qubits.
$\eta$ of these go on straight to the output, the ($\eta$+1)$^{th}$ qubit is
replaced by the parity of the ($\eta$+1) input qubits.
\end{enumerate}

The above components provide the blocks that can be used to implement the
operations $U,\ R_{s},\ U^{\dagger}\ \&\ R_{t}$ and thus the transformation
$UR_{s}U^{\dagger}R_{t}U.$

\subsubsection{ Working}

\begin{enumerate}
\item The input register is initialized with all $\left(  \eta+1\right)  $
qubits in the 0 state, one of these is the ancilla qubit. These are sent to
the modulator which flips certain qubits depending on the message to be
transmitted. The (first) parity generator computes a parity qubit and then
transmits the ($\eta$+1) qubits through the channel. All this is the
transformation $U.$

\item At the receiving end, the other parity generator computes the parity of
the ($\eta$+1) qubits and then sends the first $\eta$ of these into the output
register and the ($\eta$+1)$^{th}$ (parity) qubit into the control signal of
the output register. This is the $R_{t}$ phase shift.

\item $U^{^{\dag}}$ follows by propagating the qubits backward all the way
from the output register, through the parity generator, channel, parity
generator, modulator, all the way to the input register.

\item The input register conditionally shifts the phase if all qubits are in
the 0 state thus implementing $R_{0}$.

\item Finally the signal is propagated from the input register to the output
register again as in step 1 (this constitutes application of the last $U$ in
the transformation \underline{$\emph{U}$}$R_{s}U^{^{\dagger}}R_{t}U\left\vert
s\right\rangle $)
\end{enumerate}

Note that when classical information is being transmitted, one parity bit
would normally provide the means just to detect single bit-flip errors. The
quantum nature of the scheme enables us to \textit{correct} the error without
using any additional qubits.

\section{\bigskip Conclusion}

The variant of quantum searching discussed in this paper supplements the
original search algorithm by providing a scheme that permits a fixed point and
hence moves towards a target state in a more directed way. This new scheme has
already led to a robust quantum scheme for quantum searching that is within a
constant factor of the most efficient possible. This will be discussed in
detail elsewhere \cite{pi_3_srch},\cite{tath}.

Also it naturally leads to schemes for error correction. This paper mentions
an elementary example; a comprehensive scheme is given in \cite{reich},
\ where it is shown how to eliminate errors module by module. Other schemes
are under development.

One missing aspect is a simple physical explanation of \ how this scheme
actually works. Why does changing the $\pi$ \ phase shift in amplitude
amplification to a $\frac{\pi}{3}$ \ phase shift, convert the algorithm into
something so different? It would be insightful to have an explanation similar
to those for amplitude amplification .\bigskip\ This will be discussed further
in \cite{tath}, \cite{pi_3_srch}.

\end{document}